\documentclass[preprint,3p,twocolumn]{elsarticle}
\usepackage{atlasphysics}
\usepackage{epsfig}
\usepackage{lineno} 
\usepackage{natbib}
\usepackage{dcolumn}
\usepackage{subfigure}
\usepackage{hyperref}
\usepackage{xcolor}
\usepackage{booktabs}
\usepackage{amssymb}
\usepackage{setspace}
\usepackage{breakurl}

\usepackage{preprintcover}  
\PreprintCoverPaperTitle{Search for pair production of heavy top-like quarks decaying to a high-$\pt$ $W$ boson and a $b$ quark in the lepton plus jets final state at $\sqrt{s}=7\tev$ with the ATLAS detector}  
\PreprintIdNumber{CERN-PH-EP-2012-258}  
\PreprintCoverAbstract{
A search is presented for production of a heavy up-type quark ($t^\prime$) together with its antiparticle, assuming a significant branching ratio for 
subsequent decay into a $W$ boson and a $b$ quark. The search is
based on 4.7~fb$^{-1}$ of $pp$ collisions at $\sqrt{s}=7\tev$ recorded in 2011 
with the ATLAS detector at the CERN Large Hadron Collider.
Data are analyzed in the lepton+jets final state, characterized by a high-transverse-momentum isolated electron or muon, large missing
transverse momentum and at least three jets.
The analysis strategy relies on the substantial boost of the $W$ bosons  in  the $t^\prime \bar{t^\prime}$ signal 
when $m_{t^\prime}\gtrsim 400\gev$.
No significant excess of events above the Standard Model expectation is observed and the result of the search is interpreted in the
context of fourth-generation and vector-like quark models. Under the assumption of 
a branching ratio $BR(t^\prime \to Wb)=1$, a  fourth-generation
$t^\prime$ quark with mass lower than $656\gev$ is excluded at 95\% confidence level. 
In addition, in light of the recent discovery of a new boson of mass $\sim 126\gev$ at the LHC, 
upper limits are derived in the two-dimensional plane of $BR(t^\prime \to Wb)$ versus $BR(t^\prime \to Ht)$,  
where $H$ is the Standard Model Higgs boson, for vector-like quarks of various masses.}  
\PreprintJournalName{Physics Letters B}  

\journal{Physics Letters B}
\biboptions{square,comma,numbers,sort&compress}
\begin{document}

\begin{frontmatter}

\title{Search for pair production of heavy top-like quarks decaying to a high-$\pt$ $W$ boson and a $b$ quark in the lepton plus jets final state at $\sqrt{s}=7\tev$ with the ATLAS detector}  

\author{ATLAS Collaboration}

\begin{abstract}
A search is presented for production of a heavy up-type quark ($t^\prime$) together with its antiparticle, assuming a significant branching ratio for 
subsequent decay into a $W$ boson and a $b$ quark. The search is
based on 4.7~fb$^{-1}$ of $pp$ collisions at $\sqrt{s}=7\tev$ recorded in 2011 
with the ATLAS detector at the CERN Large Hadron Collider.
Data are analyzed in the lepton+jets final state, characterized by a high-transverse-momentum isolated electron or muon, large missing
transverse momentum and at least three jets.
The analysis strategy relies on the substantial boost of the $W$ bosons  in  the $t^\prime \bar{t^\prime}$ signal 
when $m_{t^\prime}\gtrsim 400\gev$.
No significant excess of events above the Standard Model expectation is observed and the result of the search is interpreted in the
context of fourth-generation and vector-like quark models. Under the assumption of 
a branching ratio $BR(t^\prime \to Wb)=1$, a  fourth-generation
$t^\prime$ quark with mass lower than $656\gev$ is excluded at 95\% confidence level. 
In addition, in light of the recent discovery of a new boson of mass $\sim 126\gev$ at the LHC, 
upper limits are derived in the two-dimensional plane of $BR(t^\prime \to Wb)$ versus $BR(t^\prime \to Ht)$,  
where $H$ is the Standard Model Higgs boson, for vector-like quarks of various masses.
\end{abstract}

\end{frontmatter}

\section{Introduction}

Since the discovery of the top quark~\cite{topdisc_cdf,topdisc_d0},  which completed the third generation of fundamental fermions in the
quark sector of the Standard Model (SM) of particle physics, searches for heavier quarks have been of particular interest in high-energy physics
research.  These quarks are often present in new physics models aimed at solving some of the limitations of the SM. 

One possibility is the addition of a fourth generation of heavy chiral fermions~\cite{holdom,qgupdate}, which can provide new
sources of CP violation that could explain the matter-antimatter asymmetry in the universe. 
The new weak-isospin doublet contains 
 heavy  up-type ($t^\prime$) and down-type ($b^\prime$) quarks that mix with the lighter quarks via an extended CKM matrix. 
In order to be consistent with precision electroweak data, a relatively small mass splitting between the new quarks is required~\cite{maltoni4g}. 
Assuming that $m_{t^\prime}-m_{b^\prime}<m_W$, 
where $m_W$ is the $W$ boson mass, the $t^\prime$ quark decays predominantly to a $W$ boson and a down-type quark 
$q$ ($q=d,s,b$). 
Based on the mixing pattern of the known quarks, it is natural to expect that this quark would be dominantly 
a $b$ quark, which has motivated the assumption of $BR(t^\prime \to Wb)=1$ in most experimental searches.

Another possibility is the addition of weak-isospin singlets, doublets or triplets of vector-like quarks~\cite{jaas}, 
defined as quarks for which both chiralities have the same transformation properties under
the electroweak group $SU(2) \times U(1)$. Vector-like quarks appear in many extensions 
of the SM such as little Higgs or extra-dimensional models.
In these models, a top-partner quark, for simplicity referred to here as $t^\prime$, 
often plays a key role in canceling the quadratic divergences in the Higgs boson mass induced by radiative corrections involving the top quark.
Vector-like quarks can mix preferentially with third-generation quarks, as the mixing is proportional to the mass of the SM quark~\cite{Aguila}, 
and they present a richer phenomenology than chiral quarks in fourth-generation models. 
In particular, a vector-like $t^\prime$ quark has {\em a priori} three possible decay modes, $t^\prime \to Wb$, $t^\prime \to Zt$, and $t^\prime \to Ht$,
with branching ratios that vary as a function of  $m_{t^\prime}$ and depend on the weak-isospin quantum number of the $t^\prime$ quark.
While all three decay modes can be sizable for a weak-isospin singlet, decays to only $Zt$ and $Ht$ are most
natural for a doublet. In the case of a triplet, the $t^\prime$ quark can decay either as a singlet or a doublet depending on its hypercharge.

The large centre-of-mass energy ($\sqrt{s}$) and integrated luminosity in proton-proton ($pp$) collisions produced at the 
CERN Large Hadron Collider (LHC) offer a unique opportunity to probe these models. 
At the LHC,  these new heavy quarks would be produced predominantly in pairs via the strong 
interaction for masses below $O(1\tev)$~\cite{jaas}, with sizable cross sections and clean experimental signatures. For higher masses, single production mediated by the electroweak interaction can potentially dominate, depending on the strength of the interaction between the $t^\prime$ quark and the weak gauge bosons.

Recent results of SM Higgs boson searches at the LHC have significantly impacted the prospects and focus of heavy-quark searches.
In particular, the observation of a new boson by the ATLAS~\cite{atlashiggs} and CMS~\cite{cmshiggs} Collaborations with a mass of $\sim$126~GeV and couplings close to those expected for the SM Higgs boson disfavors~\cite{abdellak,maltoni4g} fourth-generation models. These models predict a large increase 
in the production rate for $gg \to H$, which is in tension with searches in the $H\to WW^{(*)}$ and $H\to ZZ^{(*)}$ decay 
channels~\cite{atlasfourgenhiggs,cmsfourgenhiggs}.  
These results severely constrain perturbative fourth-generation models, although they may not completely exclude them yet.
For example, it has been pointed out that a fourth family of fermions can substantially modify the Higgs boson partial decay 
widths~\cite{fit4g} and various scenarios may still remain viable \cite{sasha,maltoni4g}.
At the same time, the observation of this new boson raises the level of interest
for vector-like quark searches, as $t^\prime \to Ht$ and $b^\prime \to Hb$ decays 
now have completely specified final states which offer an 
exciting opportunity for discovery of new heavy quarks.

In this Letter a search is presented for $t^\prime \bar{t^\prime}$ production using $pp$ collision data at $\sqrt{s}=7\tev$ 
collected with the ATLAS detector. The search is optimized for $t^\prime$ quark decays with large branching ratio to $Wb$. 
The lepton+jets final state signature, where one of the $W$ bosons decays leptonically and the other hadronically, 
is considered.
The most recent search by the ATLAS Collaboration in this
final state~\cite{atlas_1fb_ljets} was based on 1.04~fb$^{-1}$ of $pp$ collisions at $\sqrt{s}=7\tev$ and, under the assumption of $BR(t^\prime \to Wb)=1$, 
excluded the existence of a $t^\prime$ quark with a mass below $404\gev$ at 95\% confidence level (CL). 
A more stringent lower 95\% CL limit of $m_{t^\prime}>570\gev$~\cite{cms_new} was obtained by the CMS Collaboration
using 5.0~fb$^{-1}$ of data at $\sqrt{s}=7\tev$.
Searches have also been performed exploiting the dilepton signature resulting from the leptonic decay of both $W$ bosons.
A search by the ATLAS Collaboration in the dilepton final state using 1.04~fb$^{-1}$ of data at $\sqrt{s}=7\tev$ obtained a lower 95\% CL limit of 
$m_{t^\prime}>350\gev$~\cite{atlas_1fb_dil}. This search did not attempt to identify the flavor of the jets, making a more relaxed assumption of 
$BR(t^\prime \to Wq)=1$,  where $q$ could be any down-type SM quark.
A 95\% CL limit of $m_{t^\prime}>557\gev$~\cite{cms_5fb}, assuming $BR(t^\prime \to Wb)=1$, was obtained by the CMS Collaboration
using 5.0~fb$^{-1}$ of data at $\sqrt{s}=7\tev$.

In comparison with the previous result by the ATLAS Collaboration in the lepton+jets final state~\cite{atlas_1fb_ljets}, 
the search presented in this Letter uses almost a factor of five more data and has
revisited the overall strategy, as advocated in Refs.~\cite{holdom_1, holdom_2, holdom_3}, to take advantage of the kinematic differences 
that exist between top quark and  $t^\prime$ quark decays when $m_{t^\prime}\gtrsim 400\gev$.
In particular, the hadronically-decaying $W$ boson can be reconstructed as a single isolated jet when it is 
sufficiently boosted, leading to a significantly improved sensitivity in comparison to previous searches. In addition, the result of this
search is interpreted more generically in the context of vector-like quark models where $BR(t^\prime \to Wb)$ can be substantially smaller than unity. 
In this case the additional signals, other than $t^\prime\bar{t^\prime}\to WbWb$, contribute to the signal acceptance and are accounted for in the analysis.

\section{ATLAS detector}

The ATLAS detector~\cite{atlas_det} consists of an inner tracking system surrounded by a 
superconducting solenoid, electromagnetic and hadronic calorimeters, and a muon spectrometer.
The inner tracking system is immersed in a 2 T axial magnetic field and consists of a silicon pixel detector, a silicon microstrip detector, and a transition radiation tracker,
providing charged particle identification in the region $|\eta|<2.5$~\footnote{ATLAS uses a right-handed coordinate system with its origin at the nominal interaction point (IP) in the centre of the detector and the $z$-axis along the beam pipe. The $x$-axis points from the IP to the centre of the LHC ring, and the $y$ axis points upward. Cylindrical coordinates $(r,\phi)$ are used in the transverse $(x,y)$ plane, $\phi$ being the azimuthal angle around the beam pipe. The pseudorapidity is defined in terms of the polar angle $\theta$ as $\eta=-\ln\tan(\theta/2)$.}. 
The electromagnetic (EM) sampling calorimeter uses lead and liquid-argon.
The hadron calorimetry is based on two different detector technologies with either 
scintillator tiles or liquid argon as the active medium.  The barrel hadronic calorimeter 
consists of scintillating tiles with steel plates as the absorber material.  The endcap and 
forward hadronic calorimeters both use liquid argon, and copper or tungsten as the absorber, respectively.
The calorimeters provide coverage up to $|\eta|=4.9$.
The muon spectrometer consists of superconducting air-core toroids, a system of trigger chambers covering
the range $|\eta|<2.4$, and high-precision tracking chambers allowing muon momentum measurements in the range 
$|\eta|<2.7$. 

\section{Data sample and event preselection}

The data used in this analysis correspond to the full dataset recorded in 2011, 
and were acquired using single-electron and single-muon triggers.
The corresponding integrated luminosity is 4.7~fb$^{-1}$.

The event preselection criteria closely follow those used in recent ATLAS top quark studies~\cite{ttbar_atlas} and require 
exactly one isolated electron or muon with large transverse momentum ($\pt$), 
at least three jets among which at least one is identified as originating from a $b$ quark, 
and large missing transverse momentum ($\met$). 

Electron candidates are required to have transverse momentum $\pt >25\gev$ and $|\eta|<2.47$, excluding the transition
region ($1.37<|\eta|<1.52$) between the barrel and endcap EM calorimeters. 
Muon candidates are required to satisfy $\pt >20\gev$ and $|\eta|<2.5$. 
For leptons satisfying these $\pt$ requirements the efficiencies of the relevant single-lepton
triggers have reached their plateau values.
To reduce background from non-prompt leptons produced in semileptonic $b$- or $c$-hadron decays,
or in $\pi^\pm$/$K^\pm$ decays,
the selected leptons are required to be isolated, {\em i.e.}~to have little 
calorimetric energy or track transverse momentum around them~\cite{eperf}.
In this analysis $\tau$ leptons are not explicitly reconstructed. 
Because of the high $\pt$ threshold requirements, only a small fraction of $\tau$ leptons decaying leptonically
are reconstructed as electrons or muons, while the majority of $\tau$ leptons decaying hadronically are reconstructed as jets.

Jets are reconstructed with the anti-$k_t$ algorithm~\cite{antikt} with radius parameter $R=0.4$, from
topological clusters~\cite{topocluster} of energy deposits in the calorimeters, calibrated at the EM scale.
These jets are then calibrated to the particle (truth) level~\cite{jes} using $\pt$- and $\eta$-dependent 
correction factors derived from a combination of data and simulation.
Jets are required to have $\pt >25\gev$ and $|\eta|<2.5$.
To avoid selecting jets from other $pp$ interactions in the same bunch crossing,  at least 75\% of the sum of the $\pt$ of tracks associated
with a jet is required to come from tracks compatible with originating from
the identified hard-scatter primary vertex.
This primary vertex is chosen among the reconstructed candidates as the one with the highest 
$\sum\pt^2$ of associated tracks
and is required to have at least three tracks with $\pt>0.4\gev$.

To identify jets as originating from the hadronization of a $b$ quark ($b$ tagging), 
a continuous discriminant is produced by an algorithm~\cite{btagging} using multivariate techniques 
to combine information from the impact parameter of displaced tracks, as well as 
topological properties of secondary and tertiary decay vertices reconstructed within the jet.  
In the preselection, at least one jet is required to have a discriminant value larger than the point 
corresponding to an average efficiency 
in simulated $t\bar{t}$ events  
of $\sim 70\%$ for $b$-quark jets, 
of $\sim 20\%$ for $c$-quark jets and 
of $\sim 0.7\%$ for jets originating from light quarks ($u$, $d$, $s$) or gluons.

The $\met$ is constructed~\cite{met} from the vector sum of all calorimeter energy deposits~\footnote{Each calorimeter 
cluster/cell is considered a massless object and
is assigned the four-momentum $(E_{\rm cell}, \vec{p}_{\rm cell})$, where
$E_{\rm cell}$ is the measured energy and $\vec{p}_{\rm cell}$ is a vector
of magnitude $E_{\rm cell}$ directed from $(x,y,z)=(0,0,0)$  to the center
of the cell.} contained in topological clusters, calibrated at 
the energy scale of the associated high-$\pt$ object ({\em e.g.} jet or electron), 
and including contributions from selected muons.
Background from multi-jet production 
is suppressed by the requirement  $\met >35(20)\gev$
in the electron (muon) channel, and  
$\met +m_{\rm T}>60\gev$, where $m_{\rm T}$ 
is the transverse mass~\footnote{The transverse mass is defined by the formula
$m_{\rm T} = \sqrt{2 p^\ell_{\rm T} \met (1-\cos\Delta\phi)}$, where
$p^\ell_{\rm T}$  is the $\pt$ of the lepton and $\Delta\phi$ is the
azimuthal angle separation between the lepton and $\met$ directions.} 
of the lepton and $\met$.

\section{Background and signal modeling}

After event preselection the main background is $t\bar{t}$ production, 
with lesser contributions from the production of a $W$ boson in association with jets ($W$+jets) and multi-jet events.
Small contributions arise from single top-quark, $Z$+jets and diboson production.
Multi-jet events contribute to the selected sample mostly via the misidentification of a jet or a photon as an electron, 
or via the presence of a non-prompt lepton, {\em e.g.} from a semileptonic $b$- or $c$-hadron decay. The corresponding
yield is estimated via a data-driven method~\cite{ttbar_3pb}, which compares the number of events obtained with 
either standard or relaxed criteria for the selection of leptons.
For the $W$+jets background, the shape of the distributions of kinematic variables 
is estimated from simulation but the normalization is estimated from data 
using the predicted asymmetry between $W^+$+jets and $W^-$+jets 
production in $pp$ collisions~\cite{wasym}.
All other backgrounds, including the dominant $t\bar{t}$ background, and the signal, are estimated from simulation and normalized to their theoretical cross sections.

Simulated samples of  $t\bar{t}$ and single top-quark backgrounds (in the s-channel and for the associated production with a $W$ boson) 
are generated with {\sc MC@NLO} v4.01~\cite{mcatnlo_1,mcatnlo_2,mcatnlo_3} using the {\sc CT10} set of parton distribution functions (PDFs)~\cite{ct10}.
In the case of t-channel single-top quark production, the {\sc AcerMC} v3.8 leading-order (LO) generator~\cite{acermc} 
with the {\sc MRST LO**} PDF set~\cite{mrst} is used. 
These samples are generated assuming a top quark mass of $172.5\gev$ and are normalized to approximate next-to-next-to-LO 
(NNLO) theoretical cross sections~\cite{ttbarxs,stopxs,stopxs_2} using the {\sc MSTW2008} NNLO PDF set~\cite{mstw}. 
Samples of $W/Z$+jets events are generated with up to five additional partons using the {\sc Alpgen v2.13}~\cite{alpgen} LO generator and the 
{\sc CTEQ6L1} PDF set~\cite{cteq6}. The parton-shower and fragmentation steps are performed by 
{\sc Herwig} v6.520~\cite{herwig} in the case of {\sc MC@NLO} and  {\sc Alpgen}, and by {\sc Pythia} 6.421~\cite{py6}
in the case of {\sc AcerMC}. To avoid double-counting of partonic configurations in $W/Z$+jets events generated by both 
the matrix-element  calculation and the parton shower, a matching scheme~\cite{mlm} is employed. The $W$+jets
samples are generated separately for $W$+light jets, $Wb\bar{b}$+jets, $Wc\bar{c}$+jets, and $Wc$+jets, 
and their relative contributions
are normalized using the fraction of $b$-tagged jets in $W$+1-jet and $W$+2-jets data control samples~\cite{whf}.
The $Z$+jets background is normalized to the inclusive NNLO theoretical cross section~\cite{vjetsxs}.
The diboson backgrounds are modeled using {\sc Herwig} 
with the {\sc MRST LO**} PDF set, and 
are normalized to their NLO theoretical cross sections~\cite{dibosonxs}.
In all cases where {\sc Herwig} is used, the underlying event is simulated with 
 {\sc Jimmy} v4.31~\cite{jimmy}.

For fourth-generation $t^\prime$ quark signals, 
samples are generated with {\sc Pythia} using the {\sc CTEQ6.6} PDF set~\cite{cteq6} 
for a range of masses, $m_{t^\prime}$, from $400\gev$ to $750\gev$ in steps of $50\gev$.
For vector-like $t^\prime$ signals, samples corresponding to a singlet $t^\prime$ quark 
decaying to $Wb$, $Zt$ and $Ht$ are generated with the {\sc Protos} v2.2 LO generator~\cite{jaas,protos} 
using the {\sc CTEQ6L1} PDF set, and interfaced to {\sc Pythia} for the parton shower and fragmentation. 
The $m_{t^\prime}$ values considered range from $400\gev$ to $600\gev$ in steps of $50\gev$, and the Higgs boson mass is assumed 
to be $125\gev$. All Higgs boson decay modes are considered, with branching ratios as predicted by {\sc hdecay}~\cite{hdecay}.
For both types of signal, the samples are normalized to the approximate NNLO theoretical cross sections~\cite{ttbarxs} using the {\sc MSTW2008} NNLO PDF set.

All simulated samples include multiple $pp$ interactions and simulated events are weighted such that the distribution of the average 
number of interactions per bunch crossing agrees with data.  The simulated samples are processed through a simulation~\cite{atlas_sim}
of the detector geometry and response using {\sc Geant4}~\cite{geant}, and the same reconstruction software
as the data. Simulated events are corrected so that the physics object identification efficiencies, energy scales and energy resolutions
match those determined in data control samples, enriched in the physics objects of interest.

\section{Final selection}

\begin{table*}
\begin{center}
\begin{tabular}{lD{,}{\,\pm\,}{-1}D{,}{\,\pm\,}{-1}}
\hline
\hline
 &  \multicolumn{1}{c}{{\sl loose} selection} & \multicolumn{1}{c}{{\sl tight} selection}  \\
\hline
$t\bar{t}$  \rule{0pt}{2.6ex} \rule[-1.2ex]{0pt}{0pt} & 94 , 26 & 4.2 , 2.9 \\
$W$+jets   & 5.4 , 4.2     & 2.0 , 1.4\\
$Z$+jets   & 0.5 , 0.4     & 0.2 , 0.2 \\
Single top & 7.2 , 1.7     & 1.1 , 0.5 \\
Dibosons   & 0.1 , 0.1     & 0.04 , 0.04  \\
Multi-jet  & 5.9 , 8.4     & 3.8 , 3.2 \\
\hline
Total background  \rule{0pt}{2.6ex} \rule[-1.2ex]{0pt}{0pt} & 113 , 30 & 11.3 , 4.8 \\
Data& \multicolumn{1}{c}{$\;\;\;\,122$} & \multicolumn{1}{c}{$\;\;\;\,11$}  \\
\hline
$t^\prime\bar{t^\prime}(500\GeV)$\rule{0pt}{2.6ex} \rule[-1.2ex]{0pt}{0pt}  \\
$Wb$ : $Zt$ : $Ht$ = 1.0 : 0.0 : 0.0 \rule{0pt}{2.6ex} \rule[-1.2ex]{0pt}{0pt} & 47.4 , 6.3 & 28.2 , 3.6 \\
$Wb$ : $Zt$ : $Ht$ = 0.5 : 0.0 : 0.5 \rule{0pt}{2.6ex} \rule[-1.2ex]{0pt}{0pt} & 25.4 , 3.6 & 11.2 , 1.5 \\
\hline
\hline
\end{tabular}
\caption{\small{Number of observed events, integrated over the whole mass spectrum, compared to the SM expectation for
the combined $e$+jets and $\mu$+jets channels after the {\sl loose} and {\sl tight} selections. 
The expected signal yields assuming $m_{t^\prime}=500\gev$ for different values of 
$BR(t^\prime \to Wb)$, $BR(t^\prime \to Zt)$ and $BR(t^\prime \to Ht)$
are also shown.
 The case of $BR(t^\prime \to Wb)=1$
corresponds to a fourth-generation $t^\prime$ quark. The quoted
uncertainties include both statistical and systematic contributions.}}
\label{tab:yields}
\end{center}
\end{table*}

After preselection, 
further background suppression is achieved by applying requirements aimed at exploiting the distinct kinematic 
features of the signal. The large $t^\prime$ quark mass results in energetic $W$ bosons and $b$ quarks in the final state with large 
angular separation between them, while the decay products from the boosted $W$ bosons have small angular separation. 
The combination of these properties is very effective in suppressing the dominant
$t\bar{t}$ background since $t\bar{t}$ events with boosted $W$ boson configurations are rare, and are typically
characterized by a small angular separation between the $W$ boson and $b$ quark from the top quark decay.

\begin{figure*}
\centering
\subfigure[a][\label{fig:mwhad:type1}]{\includegraphics[width=0.40\textwidth]{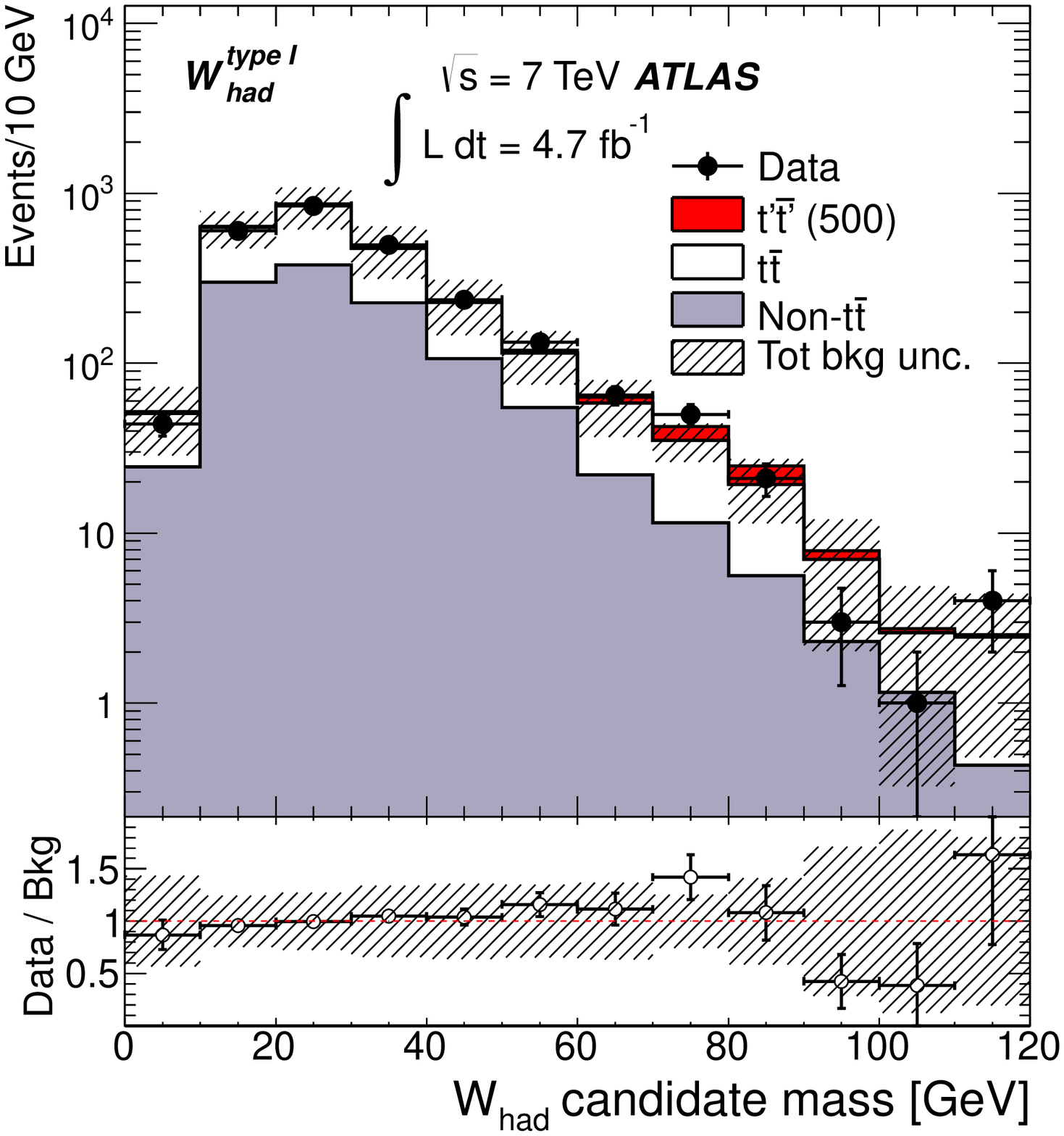}}
\subfigure[b][\label{fig:mwhad:type2}]{\includegraphics[width=0.40\textwidth]{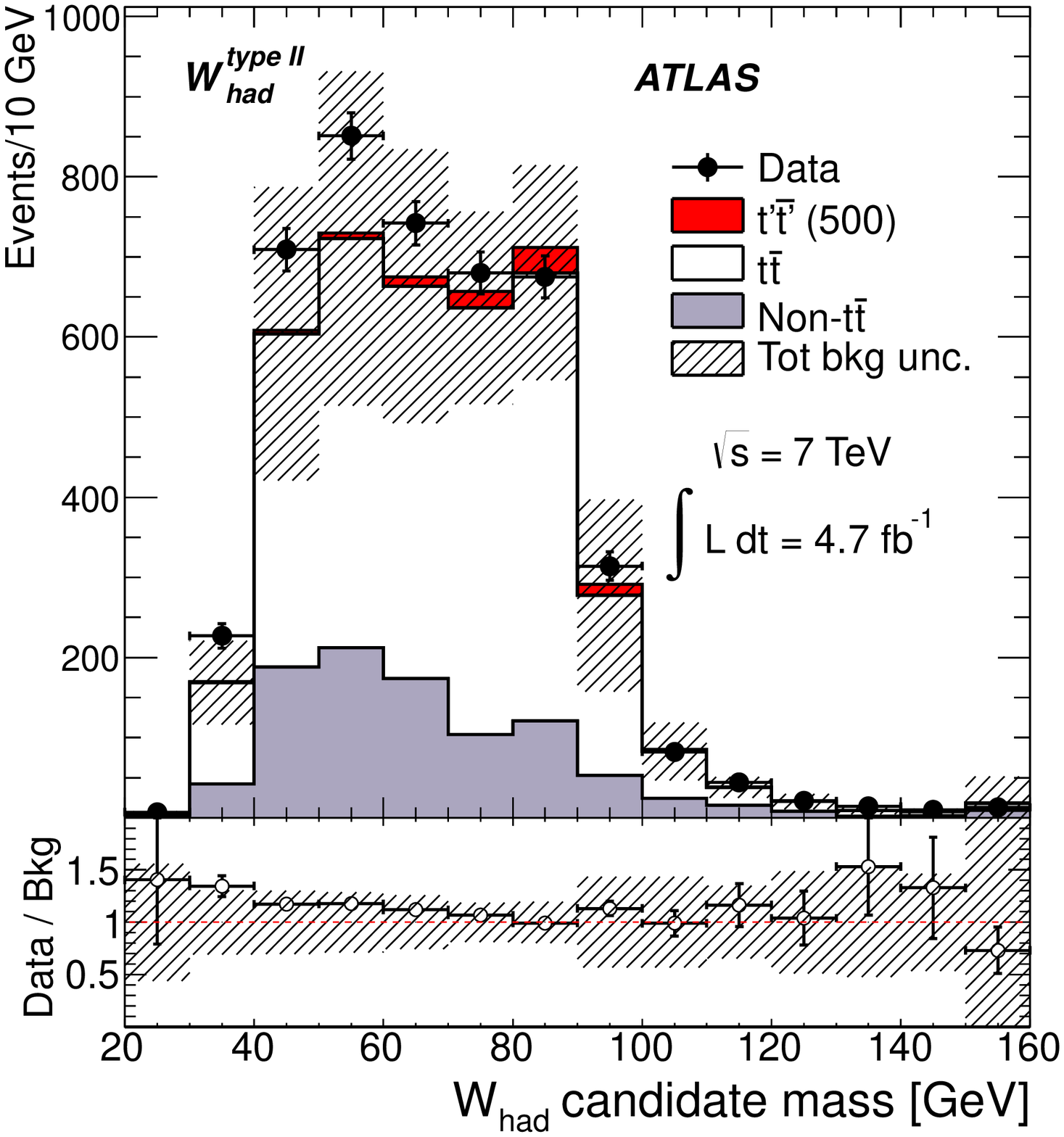}}
\caption{Distribution of the reconstructed mass for (a) $W_{\rm had}^{\rm type\;I}$ and (b) $W_{\rm had}^{\rm type\;II}$ candidates
for the combined $e$+jets and $\mu$+jets channels after preselection.  Figure (a) corresponds to events with 
 $\geq 3$ jets and $\geq 1$ $W_{\rm had}^{\rm type\;I}$ candidates, while (b) corresponds to events with 
 $\geq 4$ jets and $\geq 1$ $W_{\rm had}^{\rm type\;II}$ candidates (see text for details).
The data (solid black points) are compared to the SM prediction (stacked histograms). 
The total uncertainty on the background estimation (see Section~\ref{sec:systs} for details) is shown as a black hashed band.
The expected contribution from a fourth-generation $t^\prime$ quark with mass $m_{t^\prime}=500\gev$ is also shown (red shaded histogram), 
stacked on top of the SM background.
The last bin of each figure contains overflow events.
The lower panel shows the ratio of data to SM prediction.}
\label{fig:mwhad}
\end{figure*}

To take advantage of these properties, it is necessary to identify the 
hadronically-decaying $W$ boson ($W_{\rm had}$) as well as the $b$ jets in the event.
The candidate $b$ jets are defined as the two jets with the highest $b$-tag discriminant 
(although only one of them is explicitly required to be $b$ tagged in the event selection).
Two types of $W_{\rm had}$ candidates are defined, $W_{\rm had}^{\rm type\;I}$ and $W_{\rm had}^{\rm type\;II}$,
depending on the angular separation between their decay products. 
$W_{\rm had}^{\rm type\;I}$ is defined as a single jet with $\pt>250\gev$ 
and mass in the range of $60$--$110\gev$. 
The mass distribution for $W_{\rm had}^{\rm type\;I}$ candidates, prior to the jet mass requirement itself, is shown in Fig.~\ref{fig:mwhad:type1}.
$W_{\rm had}^{\rm type\;II}$ is defined as a dijet system with $\pt>150\gev$, angular
separation~\footnote{The angular separation is defined as $\Delta R=\sqrt{(\Delta\phi)^2 + (\Delta\eta)^2}$ where $\phi$ 
is the azimuthal angle and $\eta$ the pseudorapidity.} 
$\Delta R(j,j)<0.8$ and mass within the range of $60$--$110\gev$. If multiple pairs satisfy the above
requirements, the one with mass closest  to the nominal $W$ boson mass is chosen.
The mass distribution for $W_{\rm had}^{\rm type\;II}$ candidates, prior to the dijet mass requirement, is shown in Fig.~\ref{fig:mwhad:type2}.
In the construction of both types of $W_{\rm had}$ candidates, all selected jets except for the two candidate $b$ jets  are considered.
Small discrepancies observed between the data and the background prediction, e.g. at low $W_{\rm had}^{\rm type\;II}$ candidate invariant mass,
are not significant and covered by the systematic uncertainties.

The leptonically-decaying $W$ boson is reconstructed using the lepton and
 $\met$, identified as the neutrino $\pt$. 
Requiring that the invariant mass of the lepton--neutrino system equals the
nominal $W$ boson mass allows reconstruction of the neutrino longitudinal momentum up to a two-fold ambiguity. 
In case no real solution exists, 
the neutrino pseudorapidity is set equal to that of the lepton, since in the kinematic regime of interest for this analysis the decay products of the $W$ boson tend to be collinear.

Two final selections, {\sl loose} and {\sl tight}, are defined.
The {\sl loose} selection considers events with either $\geq 3$ jets, at least one of which is a $W_{\rm had}^{\rm type\;I}$ candidate, 
or $\geq 4$ jets, two of which combine to make at least one $W_{\rm had}^{\rm type\;II}$ candidate, and no $W_{\rm had}^{\rm type\;I}$ candidate.
The events must satisfy $\HT>750\gev$, where $\HT$ is the scalar sum of the lepton $\pt$, $\met$ and the $\pt$
of the four (or three if there are only three) highest-$\pt$ jets. The $\HT$ distribution peaks at 
$\sim 2 m_{t^\prime}$ for signal events, which makes the $\HT>750\gev$ requirement particularly efficient
for signal with $m_{t^\prime}\gtrsim 400\gev$, while rejecting a large fraction of the background. 
In addition, the highest-$\pt$ $b$-jet candidate ($b_1$) and the next-to-highest-$\pt$ $b$-jet candidate ($b_2$) are
required to have $\pt > 160\gev$ and $\pt >60\gev$, respectively.
Finally, the angular separation between the lepton and the reconstructed neutrino is required to satisfy $\Delta R(\ell,\nu)<1.4$.
The {\sl tight} selection adds  the following isolation requirements to the {\sl loose} 
selection: $\min(\Delta R(W_{\rm had}, b_{1,2}))>1.4$ and $\min(\Delta R(\ell, b_{1,2}))>1.4$, 
which are particularly effective at suppressing $t\bar{t}$ background.
Table~\ref{tab:yields} presents a summary of the background estimates for the {\sl loose} and {\sl tight} selections, as well as a
comparison of the total predicted and observed yields. 
The quoted uncertainties include both statistical and systematic contributions.
The latter are discussed 
in Section~\ref{sec:systs}.
The predicted and observed yields are in agreement within these uncertainties.

\section{Heavy-quark mass reconstruction}

\begin{figure*}
\centering
\subfigure[a][\label{fig:mreco:loose}]{\includegraphics[width=0.40\textwidth]{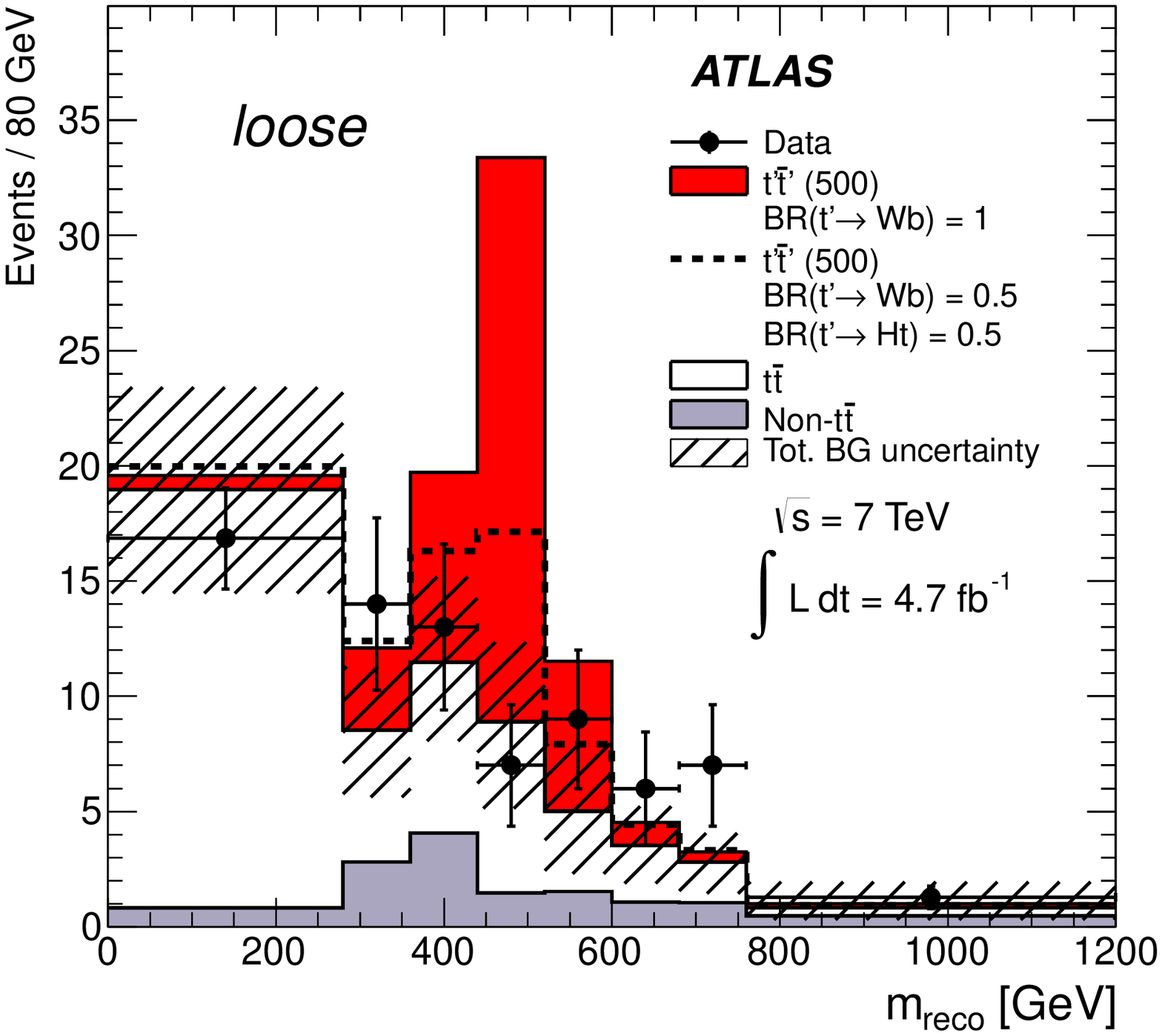}}
\subfigure[b][\label{fig:mreco:tight}]{\includegraphics[width=0.40\textwidth]{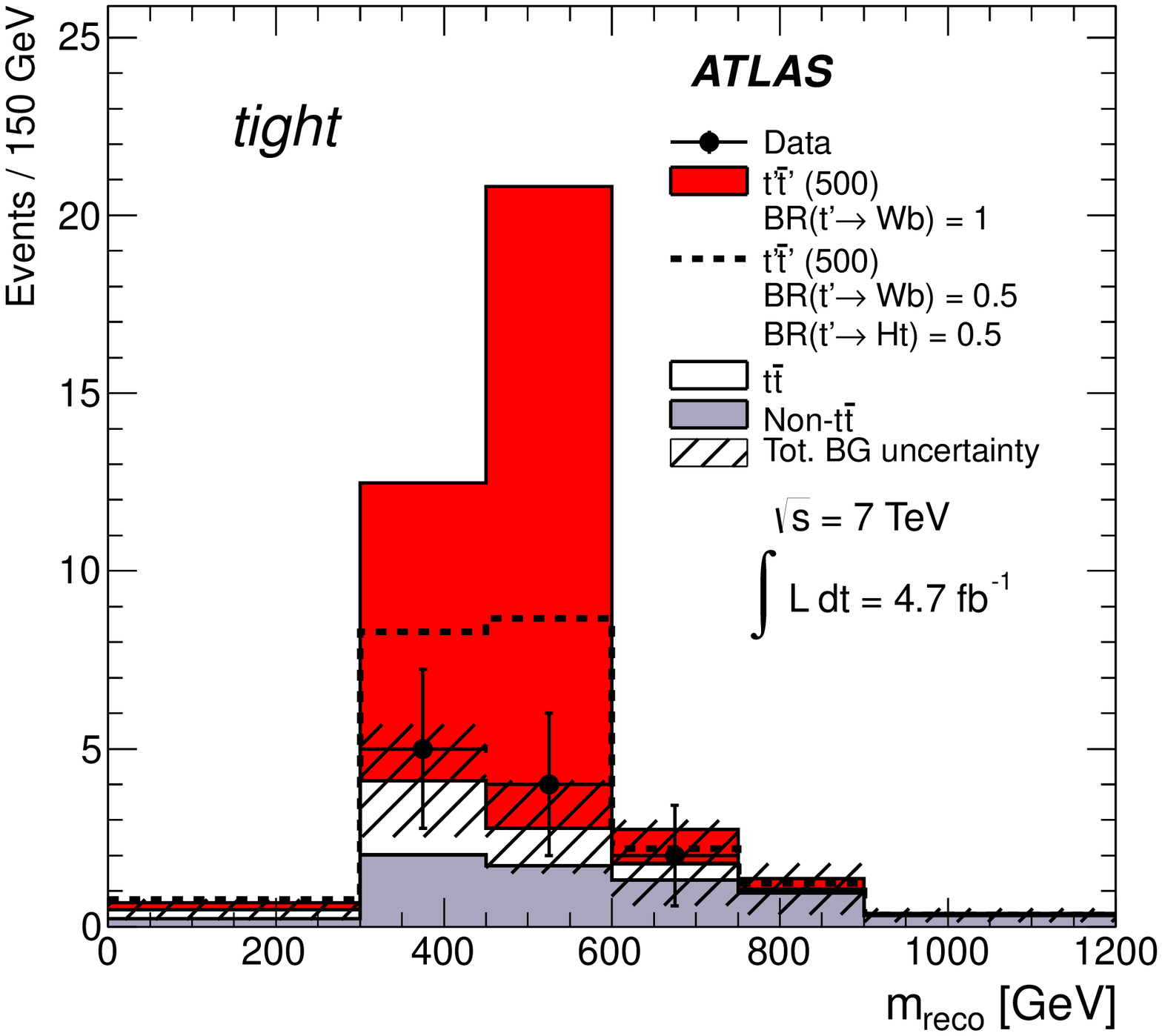}}
\caption{Distribution of $m_{\rm reco}$  for the combined $e$+jets and $\mu$+jets channels after the (a) {\sl loose} 
and (b) {\sl tight} selection. The data (solid black points) are compared to the SM prediction. 
The total uncertainty on the background estimation (see Section~\ref{sec:systs} for details) is shown as  a black hashed band.
Also shown, stacked on top of the SM background, are the expected contributions from a signal with mass $m_{t^\prime}=500\gev$ 
for the case of $BR(t^\prime \to Wb)=1$ (red shaded histogram), corresponding to a fourth-generation $t^\prime$ quark, as well
as the case of $BR(t^\prime \to Wb)=BR(t^\prime \to Ht)=0.5$ (dashed black histogram).
The overflow has been added to the last bin.}
\label{fig:mreco}
\end{figure*}

The main discriminant variable used in this search is the reconstructed heavy-quark mass ($m_{\rm reco}$), built 
from the $W_{\rm had}$ candidate and one of the two $b$-jet candidates.
The reconstruction of the leptonically-decaying $W$ boson usually yields two solutions, and there
are two possible ways to pair the $b$-jet candidates with the $W$ boson candidates to form the heavy quarks. 
Among the four possible combinations, the one yielding the smallest absolute difference between the 
two reconstructed heavy quark masses is chosen.
The resulting $m_{\rm reco}$ distributions in Fig.~\ref{fig:mreco} show that the SM background has been effectively
suppressed, and that, as is most visible for the {\sl loose} selection, good discrimination between signal and background is achieved.
The small contributions from $W$+jets, 
$Z$+jets, diboson, single-top and multi-jet events are combined into a single background source referred to as 
 non-$t\bar{t}$. It was verified {\em a priori} that the {\sl tight} selection has the best sensitivity, and it is therefore 
chosen to derive the final result for the search. The {\sl loose} selection, displaying a significant $t\bar{t}$ background
at low $m_{\rm reco}$ which is in good agreement with the expectation, provides further confidence in the 
background modeling prior to the application of $b$-jet isolation requirements in the {\sl tight} selection.

\section{Systematic uncertainties}
\label{sec:systs}

Systematic uncertainties affecting the normalization and shape of the $m_{\rm reco}$ distribution are estimated
taking into account correlations. 

Uncertainties affecting only the normalization include the integrated luminosity (3.9\%),
lepton identification and trigger efficiencies (2\%), jet identification efficiency (2\%), and cross sections for the various background
processes. The uncertainties on the theoretical cross sections for $t\bar{t}$, single-top and diboson production are  
($+9.9$/$-10.7$)\%~\cite{ttbarxs}, ($+4.7$/$-3.7$)\%~\cite{stopxs,stopxs_2}, and $\pm 5$\%~\cite{dibosonxs} respectively. 
A total uncertainty on the $W$+jets normalization of 58\% is assumed, including contributions from uncertainties on the 
$W$+4-jets cross section (48\%)~\cite{Alwall}, the heavy-flavor content measured in $W$+1,2-jets data samples (23\%)~\cite{whf},
as well as its extrapolation to higher jet multiplicities (19\%). The latter is estimated from the simulation 
where the W+heavy-flavor fractions are studied as a function of variations in the {\sc Alpgen} generator parameters.
Similarly, the $Z$+jets normalization is assigned an uncertainty of 48\% due to the dominant $Z$+4-jets contribution after final selection, 
which is evaluated at LO by {\sc Alpgen}. The multi-jet  normalization 
is assigned an uncertainty of 80\% including contributions from the limited size of the data sample (64\%) as well as the
uncertainty on the jet misidentification rate (50\%) in the data-driven prediction.

The rest of the systematic uncertainties modify both the normalization and shape of the
$m_{\rm reco}$ distribution. To indicate their magnitudes, their impact on the normalization
for the {\sl tight} selection is discussed in the following.
Among the largest uncertainties affecting the $t\bar{t}$ background are those related to modeling, 
such as (1) the choice of NLO event generator (evaluated by comparing {\sc MC@NLO} and {\sc Powheg}~\cite{powheg}),
(2) the modeling of initial- and final-state QCD radiation 
(evaluated by varying the relevant parameters in {\sc Pythia} in a range 
given by current experimental data~\cite{ttjet}), and
(3) the choice of parton-shower and fragmentation models (based on the comparison of {\sc Herwig} and {\sc Pythia}). 
These result in $t\bar{t}$ normalization uncertainties of 55\%, 1\% and 26\%, respectively.
The uncertainty on the jet energy scale~\cite{jes}
affects the normalization of the $t^\prime\bar{t^\prime}$ signal,  $t\bar{t}$ background and non-$t\bar{t}$ backgrounds
by $\pm6$\%, ($+22$/$-25$)\%, and ($+19$/$-10$)\%, respectively.
The uncertainties due to the jet energy resolution are 2\%, 3\% and 3\%, respectively.
Uncertainties associated with the jet mass scale and resolution, affecting the selection of $W_{\rm had}^{\rm type\;I}$ candidates, 
are smaller in magnitude but are also taken into account.  
Uncertainties on the modeling of the $b$-tagging algorithms affect the identification of $b$, $c$ and light jets~\cite{btagging,ctagging,ltagging},
and collectively result in uncertainties for the $t^\prime\bar{t^\prime}$ signal, as well as the $t\bar{t}$ and non-$t\bar{t}$ backgrounds, of (5--6)\%.
Other systematic uncertainties such as those on jet reconstruction efficiency or the effect of multiple $pp$ interactions on the modeling of $\met$
have been verified to be negligible.

In summary, taking into account all systematic uncertainties discussed above, the 
total uncertainty on the normalization affecting the {\sl tight} selection
for a $t^\prime\bar{t^\prime}$ signal with $m_{t^\prime}=500\gev$, 
$t\bar{t}$ and non-$t\bar{t}$ backgrounds is 11\%, 67\% 
and 50\%, respectively.

\section{Statistical analysis}

In the absence of any significant data excess, the $m_{\rm reco}$ spectrum shown in Fig.~\ref{fig:mreco:tight}
is used to derive 95\% CL upper limits on the $t^{\prime} \bar{t^\prime}$ production cross section 
 using the $CL_{\rm s}$ method~\cite{cls,cls_2}. This method employs a log-likelihood ratio $LLR=-2\log (L_{\rm s+b}/L_{\rm b})$ as test-statistic, 
where $L_{\rm s+b}$ ($L_{\rm b}$) is a binned likelihood function (product of Poisson probabilities) to observe the data under the 
signal-plus-background (background-only) hypothesis. 
Pseudo-experiments are generated for both hypotheses, taking into account per-bin statistical fluctuations
of the total predictions according to Poisson statistics, as well as Gaussian fluctuations
describing the effect of systematic uncertainties.
The fraction of pseudo-experiments for the signal-plus-background (background-only) hypothesis with $LLR$ larger than a given threshold defines $CL_{\rm s+b}$ ($CL_{\rm b}$). 
Such threshold is set to the observed (median) $LLR$ for the observed (expected) limit. 
Signal cross sections for which $CL_{\rm s}=CL_{\rm s+b}/CL_{\rm b}<0.05$ are deemed to be excluded at 95\% CL.
Dividing by $CL_{\rm b}$ avoids the possibility of mistakenly excluding a small signal due to a downward fluctuation of the background.

\section{Results}

\begin{figure}
\centering
\includegraphics[width=0.48\textwidth]{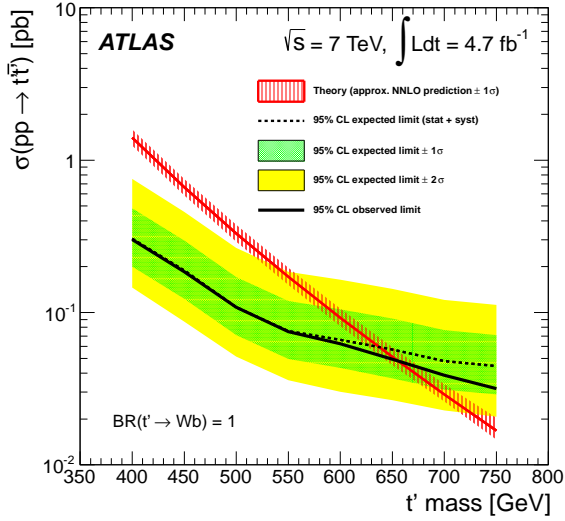}
\caption{Observed (solid line) and expected (dashed line) 95\% CL upper limits on the $t^\prime \bar{t^\prime}$ cross section 
as a function of the $t^{\prime}$ quark mass. The surrounding shaded bands correspond to the $\pm1$ and $\pm2$ standard deviations around the expected limit. 
The thin red line and band show the theoretical prediction and its $\pm1$ standard deviation uncertainty.}
\label{fig:limits}
\end{figure}

The resulting observed and expected upper limits on the $t^{\prime} \bar{t^\prime}$ production cross section are 
shown in Fig.~\ref{fig:limits} as a function of $m_{t^\prime}$, and compared to the theoretical prediction, 
assuming $BR(t^\prime \rightarrow Wb)=1$.  The total uncertainty on the theoretical cross section~\cite{ttbarxs} includes the
contributions from scale variations and PDF uncertainties. An observed (expected) 95\%  CL  limit 
$m_{t^\prime}>656\,(638)\gev$ is obtained for the central value of the theoretical cross section.
This represents the most stringent limit to date on the mass of a fourth-generation $t^\prime$ quark decaying exclusively into a $W$ boson and a $b$ quark.
This limit is also applicable to a down-type vector-like quark with electric charge of
$-4/3$ and decaying into a $W$ boson and a $b$ quark~\cite{jaas}.

\begin{figure}
\centering
\includegraphics[width=0.48\textwidth]{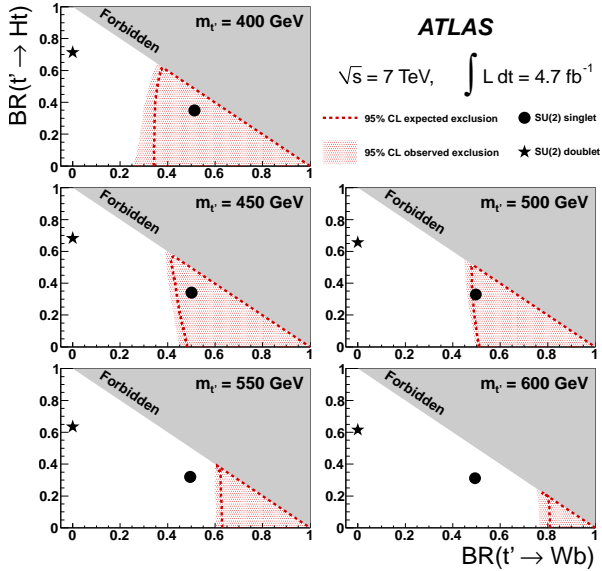}
\caption{Observed (red filled area) and expected (red dashed line) 95\% CL exclusion in the plane of
$BR(t^\prime \to Wb)$ versus $BR(t^\prime \to Ht)$, for different values of the vector-like $t^\prime$ quark mass.
The grey (dark shaded) area corresponds to the unphysical region where the sum of branching ratios exceeds unity. 
The default branching ratio values from the {\sc Protos} event generator for the weak-isospin singlet and doublet cases 
are shown as plain circle and star symbols, respectively.
\label{fig:limitsVLQ}}
\end{figure}

The same analysis is used to derive exclusion limits on vector-like $t^\prime$ quark production, for different 
values of $m_{t^\prime}$ and as a function of the two branching ratios $BR(t^\prime\to W b)$ and $BR(t^\prime\to Ht)$.
The branching ratio $BR(t^\prime\to Zt)$ is fixed by $BR(t^\prime\to Zt)= 1 - BR(t^\prime\to W b) - BR(t^\prime\to Ht)$.
To probe this two-dimensional branching-ratio plane, 
the signal samples with the original branching ratios as generated by {\sc Protos} are weighted.
The resulting 95\% CL exclusion limits are shown in Fig.~\ref{fig:limitsVLQ} for 
different values of $m_{t^\prime}$. For instance, a $t^\prime$ quark with a mass of 
$550\gev$ and $BR(t^\prime\to W b)>0.63$ is excluded at $\geq 95\%$ CL, 
regardless of the value of its branching ratios to $Ht$ and $Zt$.  
All the decay modes contribute to the final sensitivity when setting limits. 
For example, assuming $m_{t^\prime}=550\gev$, 
the efficiency of the {\sl tight} selection with at least four jets is 2.67\%, 0.64\%, 0.81\%, 0.27\%, 0.24\% and 0.25\%, for 
decays to $WbWb$, $WbHt$, $WbZt$, $ZtHt$, $ZtZt$ and $HtHt$, respectively.
The default predictions from {\sc Protos}  for the weak-isospin singlet and doublet cases 
are also shown.
A weak-isospin singlet $t^\prime$ quark with 
$400 \leq m_{t^\prime}\leq 500\gev$ is excluded at $\geq 95\%$ CL.
It should be noted that since this analysis is optimized for $m_{t^\prime}\gtrsim 400\gev$ (recall the $\HT>750\gev$ requirement), 
it is not sensitive for vector-like quark scenarios where $m_{t^\prime}<400\gev$.
The doublet scenarios are shown in Fig.~\ref{fig:limitsVLQ} to illustrate the fact that this analysis has no sensitivity in these cases. 

\section{Conclusion}

The strategy followed in this search, directly exploiting the distinct
boosted signature expected in the decay of a heavy  $t^\prime$ quark, has resulted in the
most stringent limits to date on a fourth-generation $t^\prime$ quark. 
This approach shows great promise for improved sensitivity in future LHC 
searches at higher centre-of-mass energy and integrated luminosity.
This search is also interpreted more generically in the 
context of vector-like quark models,  resulting in the first quasi-model-independent
exclusions in the two-dimensional plane of $BR(t^\prime\to W b)$ versus $BR(t^\prime\to Ht)$,
for different values of the $t^\prime$ quark mass.

\section*{Acknowledgements}

We thank CERN for the very successful operation of the LHC, as well as the
support staff from our institutions without whom ATLAS could not be
operated efficiently.

We acknowledge the support of ANPCyT, Argentina; YerPhI, Armenia; ARC,
Australia; BMWF and FWF, Austria; ANAS, Azerbaijan; SSTC, Belarus; CNPq and FAPESP,
Brazil; NSERC, NRC and CFI, Canada; CERN; CONICYT, Chile; CAS, MOST and NSFC,
China; COLCIENCIAS, Colombia; MSMT CR, MPO CR and VSC CR, Czech Republic;
DNRF, DNSRC and Lundbeck Foundation, Denmark; EPLANET and ERC, European Union;
IN2P3-CNRS, CEA-DSM/IRFU, France; GNSF, Georgia; BMBF, DFG, HGF, MPG and AvH
Foundation, Germany; GSRT, Greece; ISF, MINERVA, GIF, DIP and Benoziyo Center,
Israel; INFN, Italy; MEXT and JSPS, Japan; CNRST, Morocco; FOM and NWO,
Netherlands; BRF and RCN, Norway; MNiSW, Poland; GRICES and FCT, Portugal; MERYS
(MECTS), Romania; MES of Russia and ROSATOM, Russian Federation; JINR; MSTD,
Serbia; MSSR, Slovakia; ARRS and MVZT, Slovenia; DST/NRF, South Africa;
MICINN, Spain; SRC and Wallenberg Foundation, Sweden; SER, SNSF and Cantons of
Bern and Geneva, Switzerland; NSC, Taiwan; TAEK, Turkey; STFC, the Royal
Society and Leverhulme Trust, United Kingdom; DOE and NSF, United States of
America.

The crucial computing support from all WLCG partners is acknowledged
gratefully, in particular from CERN and the ATLAS Tier-1 facilities at
TRIUMF (Canada), NDGF (Denmark, Norway, Sweden), CC-IN2P3 (France),
KIT/GridKA (Germany), INFN-CNAF (Italy), NL-T1 (Netherlands), PIC (Spain),
ASGC (Taiwan), RAL (UK) and BNL (USA) and in the Tier-2 facilities
worldwide.

\bibliographystyle{model1-num-names}
\bibliography{EXOT-2012-07}

\onecolumn 
\clearpage
\begin{flushleft}
{\Large The ATLAS Collaboration}

\bigskip

G.~Aad$^{\rm 48}$,
T.~Abajyan$^{\rm 21}$,
B.~Abbott$^{\rm 111}$,
J.~Abdallah$^{\rm 12}$,
S.~Abdel~Khalek$^{\rm 115}$,
A.A.~Abdelalim$^{\rm 49}$,
O.~Abdinov$^{\rm 11}$,
R.~Aben$^{\rm 105}$,
B.~Abi$^{\rm 112}$,
M.~Abolins$^{\rm 88}$
% ...
\bigskip

% ...
$^{11}$ Institute of Physics, Azerbaijan Academy of Sciences, Baku, Azerbaijan\\
$^{12}$ Institut de F\'{i}sica d'Altes Energies and Departament de F\'{i}sica de la Universitat Aut\`{o}noma de Barcelona and ICREA, Barcelona, Spain\\
% ...
$^{21}$ Physikalisches Institut, University of Bonn, Bonn, Germany\\
% ...
$^{48}$ Fakult\"{a}t f\"{u}r Mathematik und Physik, Albert-Ludwigs-Universit\"{a}t, Freiburg, Germany\\
$^{49}$ Section de Physique, Universit\'{e} de Gen\`{e}ve, Geneva, Switzerland\\
% ...
$^{88}$ Department of Physics and Astronomy, Michigan State University, East Lansing MI, United States of America\\
% ...
$^{105}$ Nikhef National Institute for Subatomic Physics and University of Amsterdam, Amsterdam, Netherlands\\
% ...
$^{111}$ Homer L. Dodge Department of Physics and Astronomy, University of Oklahoma, Norman OK, United States of America\\
$^{112}$ Department of Physics, Oklahoma State University, Stillwater OK, United States of America\\
% ...
$^{115}$ LAL, Universit\'{e} Paris-Sud and CNRS/IN2P3, Orsay, France\\
% ...
\end{flushleft}


\end{document}